\documentclass[twocolumn]{jpsj3}
\usepackage{amsmath,amsthm,amssymb}

\title{
First Order Bipolaronic Transition at Finite Temperature in the Holstein Model
}

\author{Takahiro \textsc{Fuse}
\thanks{E-mail address: fuse@phys.sc.niigata-u.ac.jp (T. Fuse).} 
and 
Yoshiaki \textsc{\=Ono}}

\inst{
{\it Department of Physics, Niigata University, Ikarashi, Niigata 950-2181, Japan}
}

\abst{
We investigate the Holstein model by using the dynamical mean-field theory combined with the exact diagonalization method. Below a critical temperature $T_\mathrm{cr}$, a coexistence of the polaronic and the bipolaronic solutions is found for the same value of the electron-phonon coupling $g$ in the range $g_{c1}(T)<g<g_{c2}(T)$. In the coexistence region, the system shows a first order phase transition from the bipolaronic to the polaronic states as $T$ decreases at $T=T_p(<T_\mathrm{cr})$, where the double occupancy and the lattice fluctuation together with the anharmonicity of the effective ion potential change discontinuously without any symmetry breaking. The obtained bipolaronic transition seems to be consistent with the rattling transition in the $\beta$-pyrochlore oxide KOs$_2$O$_6$. 
}

\kword{polarons, bipolarons, bipolaronic transition, Holstein model, rattling, $\beta$-pyrochlore oxide}

\begin{document}
\maketitle

%\section{INTRODUCTION}

The electron-phonon interaction in metallic systems has been extensively studied for many years because it plays important roles in various materials such as A15 compounds\cite{C.Yu}, alkali-doped fullerides\cite{O.Gunnarsson}, magnesium diboride\cite{A.Liu} and manganites\cite{M.Imada}. 
Recently, another class of electron-phonon systems in cage structure compounds has attracted much interest, where the ion surrounded by an oversized cage shows large amplitude local vibrations called rattling\cite{G.Slack,T.Goto}. 
The superconducting $\beta$-pyrochlore oxide KOs$_2$O$_6$ with $T_c= 9.6$K shows a remarkable first order phase transition without symmetry change at $T_p= 7.5$K which is almost independent of the external magnetic field \cite{Z.Hiroi2005}. Dahm and Ueda \cite{T.Dahm} have revealed that rattling phonons with large  anharmonicity play crucial roles for the anomalous temperature dependence of the NMR relaxation rate and the resistivity observed above $T_p$. Therefore, it is important to elucidate the effect of the electron-phonon interaction on the rattling transition at $T_p$ \cite{K.Hattori} on the basis of fundamental models such as the Hubbard-Holstein model. 

Recently, the Hubbard-Holstein model has been intensively investigated using the dynamical mean-field theory (DMFT)\cite{A.Georges} which is exact in inifinite dimensions and is expected to be a good approximation in three dimensions. At zero temperature, a transition from the correlated metal to the Mott insulator (Mott transition) takes place when the on-site Coulomb interaction $U$ is dominant, while a transition from the polaronic metal to the localized bipolaronic state (bipolaronic transition) takes place when the electron-phonon interaction $g$ is dominant\cite{W.Koller,G.Jeon}. 
% The Mott transition is always found to be second order. 
The Mott transition at half-filling is always found to be second order. 
On the other hand, the bipolaronic transition is found to be first order for large values of $U$, while it is second order for small values of $U$\cite{W.Koller,P.Paci}.

%%%%%%%%%%%%%%%%%%%%%%%%%%%%%%%%%%%%%%%%%%%%%%%%%%%%%%%%%%%%%%%%%%%%
\begin{figure}[b]
%h=here, t=top, b=bottom, p=separate figure page
\begin{center}\leavevmode
\rotatebox{0}{\includegraphics[width=64mm,angle=0]{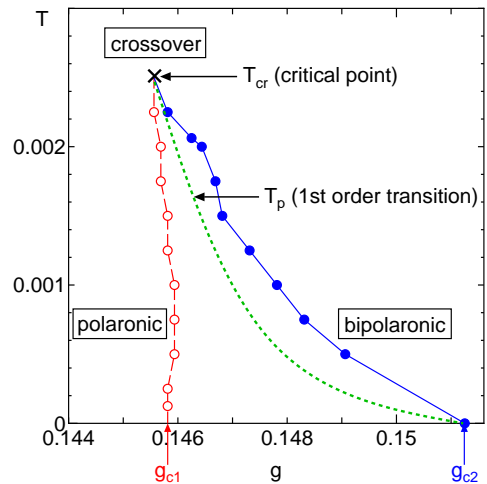}}
%\vspace{-2mm}
\caption{(Color online) 
The $g-T$ phase diagram of the polaronic and bipolaronic states. In the coexistence region $g_{c1}<g<g_{c2}$, both the polaronic and bipolaronic solutions exist. 
The dotted line shows a possible first order transition temperature $T_p$ and the cross indicates a critical point at $T=T_\mathrm{cr}$ above which the crossover between the polaronic and bipolaronic states is observed. 
\label{fig:g-T_pd2}
}
\end{center}
\end{figure}
%%%%%%%%%%%%%%%%%%%%%%%%%%%%%%%%%%%%%%%%%%%%%%%%%%%%%%%%%%%%%%%%%%%%

As for the finite temperature, we have investigated the bipolaronic transition in the Hubbard-Holstein model using the DMFT and found a coexistence of the polaronic and the bipolaronic solutions for the same value of $g$ in the range $g_{c1}(T)<g<g_{c2}(T)$\cite{T.Fuse_M2s} (see also Fig. \ref{fig:g-T_pd2} for $U=0$). In the coexistence region, the first-order bipolaronic transition takes place at $T_p$ below a critical temperature $T_\mathrm{cr}$ above which the smooth crossover is observed instead of the transition. For $T>T_p$, the system is in the bipolaronic state with large lattice fluctuation, while,  for $T<T_p$, it is in the polaronic state  with relatively small lattice fluctuation. 
 The effective potential for oscillating ions is double-well type for $T>T_p$, while it is  single-well type for $T<T_p$. 
When $U$ decreases, the discontinuities in physical quantities at $T_p$ decrease together with decreasing $T_\mathrm{cr}$ and the difference $|g_{c1}-g_{c2}|$.  Especially, in the Holstein model with $U=0$, both of $T_\mathrm{cr}$ and $|g_{c1}-g_{c2}|$ become very small resulting in numerical difficulties in detecting the coexistence region even for $T=0$\cite{D.Meyer}. 
In this paper, we perform the detailed calculations at low temperatures down to $1/10^4$ of the bandwidth to obtain the phase diagrm of the polaronic and the bipolaronic states in the Holstein model as shown in Fig. \ref{fig:g-T_pd2}.

%\section{METHOD}
The Holstein model is given by the following Hamiltonian 
%%%%%%%%%%%%%%%%%%%%%%%%%%%%%%%%%%%%% EQUATION %%%%%%%%%%%%%%%%%%%%%%%%%%%%%%%%%%%%%%
\begin{multline}
 H=\sum_{\mathbf{k}\sigma} \epsilon_\mathbf{k}c^\dag_{\mathbf{k}\sigma}c_{\mathbf{k}\sigma}
  +\omega_0 \sum_i {b_i}^\dag b_i \\
  +g\sum_i \left(b_i^\dag +b_i\right)\left(\sum_\sigma \hat{n}_{i\sigma}-1\right),
\label{eq:HolsteinHamiltonian}
\end{multline}
%%%%%%%%%%%%%%%%%%%%%%%%%%%%%%%%%%%%%%%%%%%%%%%%%%%%%%%%%%%%%%%%%%%%%%%%%%%%%%%%%%%%%
where $c^\dag_{\mathbf{k}\sigma}$ ($c^\dag_{i\sigma}$) is a creation operator for a conduction electron with wave vector $\mathbf{k}$ (site $i$) and spin $\sigma$, and $\hat{n}_{i\sigma}=c^\dag_{i\sigma}c_{i\sigma}$. 
$b_i$ is a creation operator for a phonon at site $i$, where the lattice displacement operator is given by $\hat{Q}_i=(b_i^\dag+b_i)/\sqrt{2\omega_0}$. 
$\epsilon_\mathbf{k}$, $\omega_0$ and $g$ are 
the energy for a conduction electron, 
the frequency of the Einstein phonons and 
the electron-phonon coupling strength, respectively. 

To solve the model eq.(\ref{eq:HolsteinHamiltonian}), 
we use the DMFT \cite{A.Georges} in which the model is mapped onto an effective impurity Anderson-Holstein model \cite{D.Meyer}. 
In the case of a semielliptic DOS for the bare conduction band with the bandwidth $W=1$, $\rho(\epsilon)=4\sqrt{1-4\epsilon^2}/\pi$, the local Green's function $G(i\omega_n)$ satisfies the following self-consistency condition, 
%%%%%%%%%%%%%%%%%%%%%%%%%%%%%%%%%%%%% EQUATION %%%%%%%%%%%%%%%%%%%%%%%%%%%%%%%%%%%%%%
%\begin{equation} 
$
{\cal G}_0(i\omega_n)^{-1} = i\omega_n-\mu-(W/4)^2 G(i\omega_n), 
$
%\label{eq:LGFforInfiniteDimension}
%\end{equation} 
%%%%%%%%%%%%%%%%%%%%%%%%%%%%%%%%%%%%%%%%%%%%%%%%%%%%%%%%%%%%%%%%%%%%%%%%%%%%%%%%%%%%%
where $\mu$ is the chemical potential and ${\cal G}_0(i\omega_n)$ is the bare local Green's function for the effective impurity Anderson-Holstein model with $g=0$ in an effective medium which will be determined self-consistently. 
The effective impurity Anderson-Holstein model with finite $g$ is solved by using the exact diagonalization method for a finite-size cluster to obtain $G(i\omega_n)$ at finite temperature $T>0$. 
In the present paper, we use 5 site cluster and the cutoff of phonon number is set to be 12. We note that the numerical results for $6$-site are almost the same as those for $5$-site, and the numerical results for $15$ phonons are almost the same as those for $12$ phonons. 
We concentrate our attention on the particle-hole symmetric case at half-filling with $\langle{\hat{n}_i}\rangle= \langle{\sum_\sigma\hat{n}_{i\sigma}}\rangle=1$, and we set $\omega_0=0.1$.

\begin{figure}[t]
%h=here, t=top, b=bottom, p=separate figure page
\begin{center}\leavevmode
\rotatebox{0}{\includegraphics[width=60mm,angle=0]{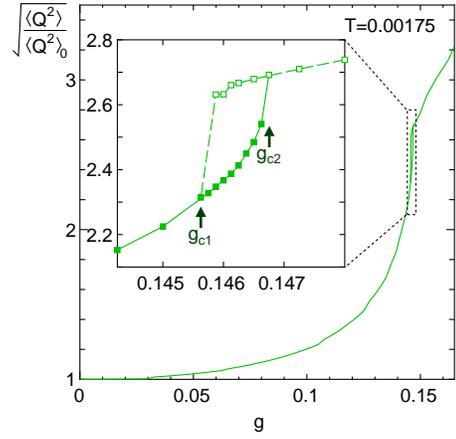}}
%\vspace{-2mm}
\caption{(Color online) 
The square root of the normalized local lattice fluctuation $\sqrt{\langle{Q^2}\rangle/\langle{Q^2}\rangle_0}$ as a function of $g$ for $T=0.00175$. 
The inset shows the magnification around the coexistence region $g_{c1}<g<g_{c2}$ where a coexistence of small $\langle{Q^2}\rangle$ (closed squares) and large $\langle{Q^2}\rangle$ (open squares) solutions is observed.  

\label{fig:g-rootQ2}
}
\end{center}
\end{figure}

The local lattice fluctuation is defined by 
$\langle{Q^2}\rangle=\langle{(\hat{Q}_i-\langle{\hat{Q}_i}\rangle)^2}\rangle$. 
Fig. \ref{fig:g-rootQ2} shows the square root of the normalized local lattice fluctuation $\sqrt{\langle{Q^2}\rangle/\langle{Q^2}\rangle _0}$ as a function of $g$ for $T=0.00175$,  where $\langle{Q^2}\rangle_0=1/2\omega_0$ is the value for the zero-point oscillation with $g=0$. 
When $g$ increases, $\langle{Q^2}\rangle$ increases gradually for small $g$, 
while it does steeply at $g\sim 0.146$, 
and then finally shows a linear increase for large $g$. 
Remarkably, a coexistence of two solutions: solutions with small and large lattice fluctuation $\langle{Q^2}\rangle$, is observed for the same value of $g$ in the range $g_{c1}<g<g_{c2}$ as shown in the inset of Fig. \ref{fig:g-rootQ2}. 
For $g<g_{c1}$ ($g>g_{c2}$), the solution with large (small) $\langle{Q^2}\rangle$ disappears and the solution with small (large) $\langle{Q^2}\rangle$ exclusively exists. 
In the coexistence region, the system shows a first order phase transition which will be discussed later. 

%%%%%%%%%%%%%%%%%%%%%%%%%%%%%%%%%%%%%%%%%%%%%%%%%%%%%%%%%%%%%%%%%%%%
%%%%%%%%%%%%%%%%%%%%%%%%%%%%%%%%%%%%%%%%%%%%%%%%%%%%%%%%%%%%%%%%%%%%

In Figs. \ref{fig:g-Z,d,cs,cc,sel,sep} (a)-(f), we plot various physical quantities as functions of $g$ around $g_{c1(2)}$ for several temperatures $T=0.001$, 0.00175 and 0.003. 
Fig. \ref{fig:g-Z,d,cs,cc,sel,sep} (a) shows the $g$ dependence of 
the square root of the normalized local lattice fluctuation $\sqrt{\langle{Q^2}\rangle/\langle{Q^2}\rangle_0}$. 
At low temperatures $T=0.001$ and 0.00175, we observe the coexistence region in the range $g_{c1}(T)<g<g_{c2}$(T), where the range for higher temperature $T=0.00175$ is smaller than that for lower temperature $T=0.001$. 
On the other hand, at high temperature $T=0.003$, the coexistence region disappears and the system shows a smooth crossover between small and large lattice fluctuationg solutions.

%%%%%%%%%%%%%%%%%%%%%%%%%%%%%%%%%%%%%%%%%%%%%%%%%%%%%%%%%%%%%%%%%%%%
\begin{figure}[t]
%h=here, t=top, b=bottom, p=separate figure page
\begin{center}\leavevmode
\rotatebox{0}{
\includegraphics[width=80mm,angle=0]{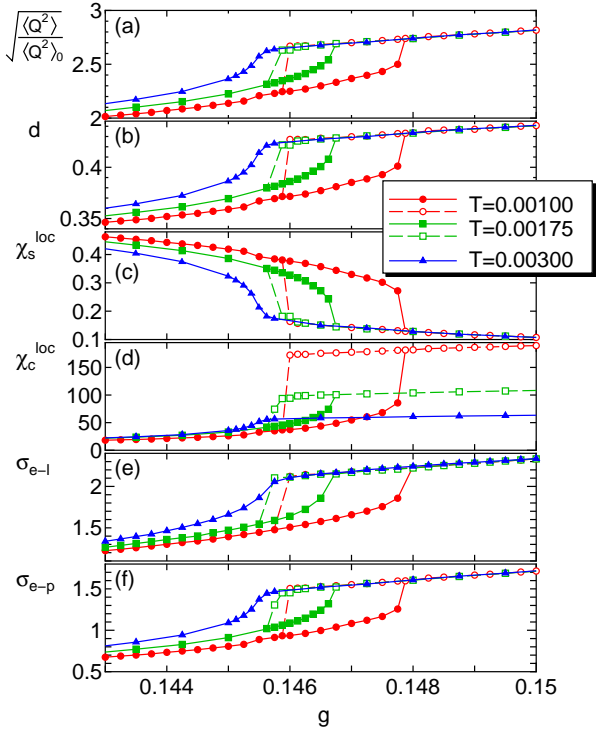}}
%\vspace{-2mm}
\caption{(Color online) 
The square root of the normalized local lattice fluctuation $\sqrt{\langle{Q^2}\rangle/\langle{Q^2}\rangle_0}$ (a), 
the double occupancy $d$ (b),
the local spin susceptibility $\chi_s^\mathrm{loc}$ (c), 
the local charge susceptibility $\chi_c^\mathrm{loc}$ (d), 
the electron-lattice cross correlation function $\sigma_{e-l}$ (e) 
and the electron-phonon density correlation function $\sigma_{e-p}$ (f) 
as functions of $g$ around the coexistence region 
for $T=0.001$, $0.00175$ and $0.003$.
\label{fig:g-Z,d,cs,cc,sel,sep}
}
\end{center}
\end{figure}
%%%%%%%%%%%%%%%%%%%%%%%%%%%%%%%%%%%%%%%%%%%%%%%%%%%%%%%%%%%%%%%%%%%%

Fig. \ref{fig:g-Z,d,cs,cc,sel,sep} (b) shows the $g$ dependence of 
the double-occupancy $d=\langle{\hat{n}_\uparrow \hat{n}_\downarrow}\rangle$. 
When $g$ increases, $d$ increases gradually for small $g$ from the noninteracting value $d=0.25$ (not shown), while it does steeply at $g\sim g_{c1(2)}$, 
and then finally shows a linear increase for large $g$. 
Similar to $\langle{Q^2}\rangle$, we observe coexistence of a small $d$ solution (polaronic state) and a large $d$ solution (bipolaronic state) in the range $g_{c1}(T)<g<g_{c2}(T)$ for low temperatures $T=0.001$ and 0.00175, while a smooth crossover between small and large $d$ solutions is observed at high temperature $T=0.003$. 
In the bipolaronic state with large $d$, the local charge fluctuation, 
$\langle{(\hat{n}_i-\langle{\hat{n}_i}\rangle)^2}\rangle=2d$, 
is enhanced together with the local lattice fluctuation $\langle{Q^2}\rangle$, while the local moment, $\langle{\mathbf{s}_i^2}\rangle=3(1-2d)/4$, is suppressed. 

To discuss the spin and charge properties more directly, we calculate the local spin susceptibility, 
$\chi_s^\mathrm{loc}=\int^\beta_0 \langle{T_\tau \hat{s}_{zi}(\tau)\hat{s}_{zi}(0)}\rangle d\tau$ with $\hat{s}_{zi}=(\hat{n}_{i\uparrow}-\hat{n}_{i\downarrow})/2$, 
and the local charge susceptibility, 
$\chi_c^\mathrm{loc}=\int^\beta_0 \langle{T_\tau \hat{n}_i(\tau)\hat{n}_i(0)\rangle} d\tau$, and plot them as functions of $g$ in Figs. \ref{fig:g-Z,d,cs,cc,sel,sep} (c) and (d). 
When $g$ increases, $\chi_s^\mathrm{loc}$ ($\chi_c^\mathrm{loc}$) decreases (increases) gradually for small $g$ (not shown), while it does steeply at $g\sim g_{c1(2)}$, and then finally shows a linear decrease (increase) for large $g$. 
The enhancement (suppression) of $\chi_c^\mathrm{loc}$ ($\chi_s^\mathrm{loc}$) due to the effect of $g$ is consistent with the local charge fluctuation (the local moment) mentioned above. We note that, $\chi_c^\mathrm{loc}$ in the bipolaronic state increases with decreasing $T$ in proportion to $1/T$ at low temperature as explicitly shown later. 

To examine the correlation between the lattice (or the phonon) and the charge fluctuation in more detail, we calculate the electron-lattice cross correlation function\cite{F.Marsiglio}, 
$\sigma_{e-l}=\langle{(b_i^\dag +b_i)(\hat{n}_i-\langle{\hat{n}_i}\rangle)}\rangle$, 
and the electron-phonon density correlation function\cite{G.Wellein}, 
$\sigma_{e-p}=\langle{ b_i^\dag b_i (\hat{n}_i-\langle{\hat{n}_i}\rangle)}\rangle$. 
As shown in Figs. \ref{fig:g-Z,d,cs,cc,sel,sep} (e) and (f), 
both $g$ dependence of $\sigma_{e-l}$ and $\sigma_{e-p}$ are very similar to those of the local lattice fluctuation and the double occupancy (i. e., the local charge fluctuation) shown in Figs. \ref{fig:g-Z,d,cs,cc,sel,sep} (a) and (b). 
Then, the strong correlation between the lattice (phonon) and the charge fluctuation due to the effects of the electron-phonon coupling $g$ is responsible for the bipolaronic state with enhanced lattice and charge fluctuations.

For various temperatures, we examine the $g$ dependence of physical quantities to obtain the coexistence region $g_{c1}(T)<g<g_{c2}(T)$ (see Fig. \ref{fig:g-Z,d,cs,cc,sel,sep} for $T=0.001$ and $0.00175$). Fig. \ref{fig:g-T_pd2} shows the $g-T$ phase diagram where the critical values $g_{c1}(T)$ and $g_{c2}(T)$ are plotted. 
When $T$ increases, $g_{c2}(T)$ decreases monotonically, 
while $g_{c1}(T)$ is almost independent of $T$, 
and then $g_{c1}(T)$ and $g_{c2}(T)$ coincide with each other at a critical temperature $T_\mathrm{cr}\sim 0.0025$. 
For $T>T_\mathrm{cr}$, the coexistence region disappears and the system shows a smooth crossover between the polaronic and bipolaronic states  (see Fig. \ref{fig:g-Z,d,cs,cc,sel,sep} for $0.003$). 

The previous DMFT studies for $T=0$\cite{D.Meyer,W.Koller} revealed that a second order phase transition from the polaronic metal to the localized bipolaronic state takes place with increasing $g$ at $g=g_{c2}$. At finite temperature below $T_\mathrm{cr}$, we find that a first order phase transition between the bipolaronic state ($T>T_p$) and the polaronic state ($T<T_p$)  takes place at a transition temperature $T_p$ in the coexistence region, similar to the case with the Mott transition observed in the Hubbard model\cite{A.Georges}. Although explicit calculations of the free energy to obtain $T_p$ have not been done so far, a possible value of $T_p$ is shown in Fig. \ref{fig:g-T_pd2}.

%%%%%%%%%%%%%%%%%%%%%%%%%%%%%%%%%%%%%%%%%%%%%%%%%%%%%%%%%%%%%%%%%%%%
\begin{figure}[t]
%h=here, t=top, b=bottom, p=separate figure page
\begin{center}\leavevmode
\rotatebox{0}{\includegraphics[width=75mm,angle=0]{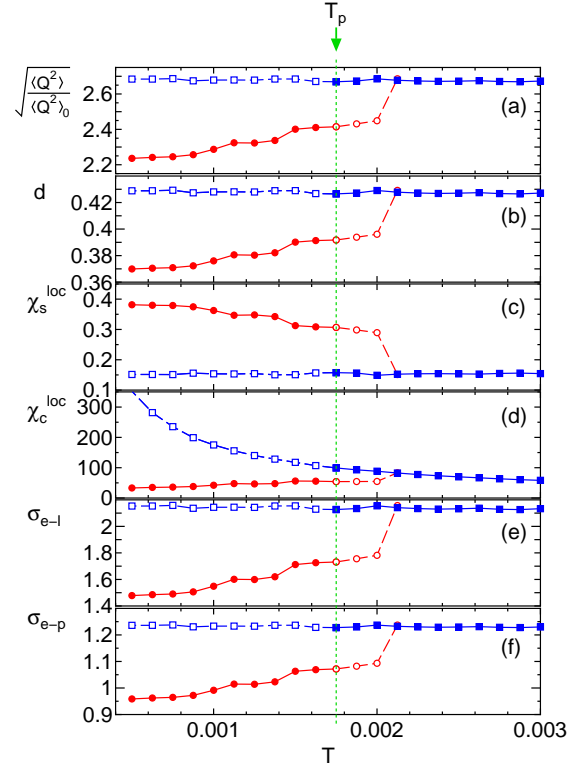}}
%\vspace{-2mm}
\caption{(Color online) 
$T$ dependence of $\sqrt{\langle{Q^2}\rangle/\langle{Q^2}\rangle_0}$, 
$d$, $\chi_s^\mathrm{loc}$, $\chi_c^\mathrm{loc}$, 
$\sigma_{e-l}$ and $\sigma_{e-p}$ for $g=0.1463$. 
The circles (squares) correspond to the polaronic (bipolaronic) solutions and the closed (open) symbols  correspond to the stable (metastable) solutions. $T_p$ is a possible value of the first order phase transition temperature. 
\label{fig:T-dep_aroundTp}
}
\end{center}
\end{figure}
%%%%%%%%%%%%%%%%%%%%%%%%%%%%%%%%%%%%%%%%%%%%%%%%%%%%%%%%%%%%%%%%%%%%

In Figs. \ref{fig:T-dep_aroundTp} (a)-(f), 
we show the temperature dependence of the physical quantities which are the same as those in Figs. \ref{fig:g-Z,d,cs,cc,sel,sep} (a)-(f) for a fixed value of $g=0.1463$. 
A coexistence of the polaronic and bipolaronic solutions is observed for low temperature $T \lesssim 0.002$, while the bipolaronic solution is exclusively observed for high temperature $T\gtrsim 0.002$. 
As shown in the $g-T$ phase diagram in Fig. \ref{fig:g-T_pd2}, it is expected that the system shows a first order phase transition between the bipolaronic state ($T>T_p$) and the polaronic state ($T<T_p$) at a transition temperature $T_p \sim 0.00175$ shown in Figs. \ref{fig:T-dep_aroundTp} (a)-(f).

In the polaronic state, all of $\langle{Q^2}\rangle$, $d$, $\chi_c^\mathrm{loc}$, $\sigma_{e-l}$ and $\sigma_{e-p}$ gradually increase with increasing $T$ together with increasing thermal excitation of phonons, while $\chi_s^\mathrm{loc}$ decreases. 
On the other hand, in the bipolaronic state, all of $\langle{Q^2}\rangle$, $d$, $\chi_s^\mathrm{loc}$, $\sigma_{e-l}$ and $\sigma_{e-p}$ are almost independent of $T$, where the quantum lattice and charge fluctuations are fully enhanced due to the strong coupling effects even for $T=0$ and then the thermal excitation  effects are relatively suppressed. 
We note that the local charge susceptibility in the bipolaronic state increases with decreasing $T$ and shows a Curie law behavior $\chi_c^\mathrm{loc} \propto 1/T$ at low temperature where the bipolarons are almost localized. This is  similar to the case with the localized spins in the Mott insulator, where the local spin susceptibility shows a Curie law behavior $\chi_s^\mathrm{loc} \propto 1/T$.

\begin{figure}[t]
%h=here, t=top, b=bottom, p=separate figure page
\begin{center}\leavevmode
\rotatebox{0}{\includegraphics[width=60mm,angle=0]{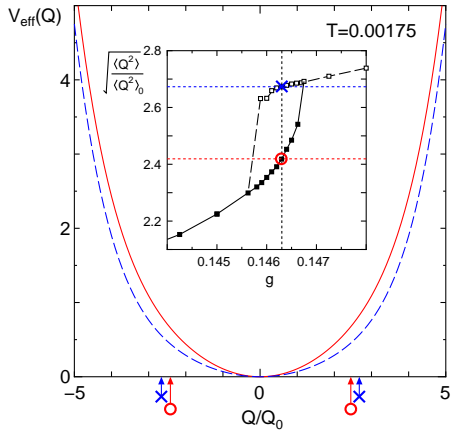}}
%\vspace{-2mm}
\caption{(Color online) 
The effective potential for oscillating ions $V_\mathrm{eff}(Q)$  at $g=0.1463$ and $T=0.00175$ as a function of the normalized displacement of the ion $Q/Q_0$ with $Q_0=1/\sqrt{2\omega_0}$. Solid (dashed) line indicates the result of the polaronic (bipolaronic) solution corresponding to the open circle (cross) in the inset. 
\label{fig:Veff_T00070}
}
\end{center}
\end{figure}

Finally, we calculate the effective potential for oscillating ions $V_\mathrm{eff}(Q)$, which is renormalized due to the effect of $g$, by using a variational wave function as previously done in refs. [\cite{S.Yotsuhashi,K.Mitsumoto}].  
Fig. \ref{fig:Veff_T00070} shows $V_\mathrm{eff}(Q)$ for the polaronic and bipolaronic solutions at the same parameters $g=0.1463$ and $T=0.00175$, where the first order phase transition between the polaronic and bipolaronic states is expected to take place as mentioned before. When $g$ increases, the harmonic term in $V_\mathrm{eff}(Q)$ decreases while the anharmonic terms increase (not shown), resulting in a largely anharmonic effective potential in the strong coupling regime. In the coexistence region, the anharmonicity for the bipolaronic state is larger than that for the polaronic state as shown in Fig. \ref{fig:Veff_T00070}. Then, the first order bipolaronic transition is accompanied by the change in the anharmonicity of the effective potential. 
It is noted that, in the Hubbard-Holstein model, $V_\mathrm{eff}(Q)$ for the bipolaronic state is found to be strongly anharmonic double-well type for large $U$\cite{T.Fuse_M2s}, but the anharmonicity decreases with decreasing $U$ together with decreasing the discontinuities in physical quantities at $T_p$, and then $V_\mathrm{eff}(Q)$ becomes largely anharmonic single-well type for $U=0$ as shown in Fig. \ref{fig:Veff_T00070}.

%%%%%%%%%%%%%%%%%%%%%%%%%%%%%%%%%%%%%%%%%%%%%%%%%%%%%%%%%%%%%%%%%%%%
%%%%%%%%%%%%%%%%%%%%%%%%%%%%%%%%%%%%%%%%%%%%%%%%%%%%%%%%%%%%%%%%%%%%

%\section{DISCUSSTION}
In summary, we have investigated the half-filled Holstein model by using the dynamical mean-field theory and found that the system shows a first order phase transition from the bipolaronic to the polaronic states as $T$ decreases at a transition temperature $T_p$ below a critical temperature $T_\mathrm{cr}$ above which the smooth crossover is observed instead of the transition. 
At $T=T_p$, we observe the discontinuous changes in various physical quantities such as  the double occupancy, the  local lattice fluctuation and the anharmonicity of the effective potential for oscillating ions, without any symmetry breaking similar to the liquid-gas transition. 

In this paper, we set the phonon frequency $\omega_0=0.1W$, where the value of the critical temperature is obtained as $T_\mathrm{cr}=0.0025W$. 
 For large $\omega_0 \gg W$, the Holstein model is known to coincide with the attractive Hubbard model with the effective on-site Coulomb interaction $U_\mathrm{eff}=-2g^2/\omega_0$. At half-filling, the attractive Hubbard model can be transformed into the repulsive Hubbard model\cite{Micnas}, where the first-order Mott transition takes place below the critical temperature $T_\mathrm{cr}=0.013W$\cite{J.Joo}. Therefore, the bipolaronic transition for $\omega_0 \gg W$ is expected to take place below the same critical temperature $T_\mathrm{cr}=0.013W$, although the physics is different from the Mott transition. 
 When $\omega_0$ decreases,  the critical temperature monotonically decreases from $T_\mathrm{cr}=0.013W$ for $\omega_0=\infty$ to $T_\mathrm{cr}\to 0$ for $\omega_0\to 0$. At the same time, the critical values of the bipolaronic transition $g_{c1,2}$ and the discontinuities in physical quantities at $T_p$ decrease with decreasing $\omega_0$. 
We note that, all of $T_\mathrm{cr}$,  $g_{c1,2}$ and the discontinuities  at $T_p$ increase with increasing $U$ as shown in our previous study for the Hubbard-Holstein model\cite{T.Fuse_M2s}.

The first order Mott transition is observed in several materials below a critical temperature $T_\mathrm{cr}$, for example, the transition metal oxide V$_2$O$_3$ with $T_\mathrm{cr}\sim 400$K\cite{D.McWhan} and the organic conductor $\kappa$-(ET)$_2$Cu[N(CN)$_2$]Cl with $T_\mathrm{cr}\sim 35$K\cite{S.Lefebvre}, where the values of $T_\mathrm{cr}$ are consistent with the corresponding values of the conduction bandwidth of orders of 1eV for V$_2$O$_3$ and 0.1eV for organics. In these materials, magnetic ordering transition temperatures are considered to be suppressed due to the frustration effect and become lower than $T_\mathrm{cr}$, resulting in the direct observations of the Mott transition. 
Then, we may also observe the bipolaronic transition when charge ordering transition temperatures are suppressed and become lower than $T_\mathrm{cr}$ in strongly coupled electron-phonon systems. 
A promising candidate is the rattling transition in KOs$_2$O$_6$. 
In fact, the observed first order transition temperature $T_p= 7.5$K is consistent with $T_\mathrm{cr}\sim 10$K predicted from the calculation with realistic parameters of the bandwidth $W\sim 3$eV\cite{R.Sainz} and the phonon frequency $\omega_0 \sim 30$meV. 

When the pressure is increased, $g/W$ ($U/W$) decreases with increasing $W$, and then the first order transition disappears as clearly understood from the $g-T$ ($U-T$) phase diagram. Such pressure induced disappearance of the first order transition is observed in KOs$_2$O$_6$\cite{J.Yamaura_PrivateC} as well as in V$_2$O$_3$\cite{D.McWhan}. 
In addition, the effects of substitution and/or randomness are known to induce the effective pressure\cite{Byczuk} and is also expected to responsible for the disappearance of the first order transition. 
Actually, the rattling transition in KOs$_2$O$_6$ depends on the quality of samples and disappears for low-purity samples\cite{Z.Hiroi2005}. 
To be more conclusive, we need further investigation on the electron-phonon systems with including the effects of cage and/or band structures together with the superconductivity.

\section*{Acknowledgments}
The authors thank T. Goto, Y. Nemoto, H. Tsunetsugu and K. Hattori for useful comments and discussions. 
This work was supported in part by a Grant-in-Aid for Scientific Research from the Ministry of Education, Culture, Sports, Science, and Technology of Japan.

\end{document}